\documentclass[seceq,preprint]{ptptex}

\usepackage{graphicx}

\newcommand{\nn}{\nonumber\\}
\newcommand{\bra}[1]{\left<#1 \right|}
\newcommand{\ket}[1]{\left| #1 \right>}

\newcommand{\rL}{{\rm L}}

\newcommand{\Q}{Q_{\rm B}}
\newcommand{\tQ}{\tilde{Q}_{\rm B}}
\newcommand{\bQ}{\bar{Q}_{\rm B}}

\preprintnumber[3cm]{
hep-th/0304261}

\title{
Gauge Fixing and Scattering Amplitudes in String\\
Field Theory around Universal Solutions
}

\author{
Tomohiko \textsc{Takahashi}$^{1,}$\footnote{E-mail:
tomo@asuka.phys.nara-wu.ac.jp} 
and Syoji \textsc{Zeze}$^{2,}$\footnote{E-mail:
zeze@sci.osaka-cu.ac.jp.}
}

\inst{
$^1$Department of Physics, Nara Women's University, Japan\\
$^2$Department of Physics, Osaka City University,
Japan
}

\abst{
We study a gauge fixed action of open string field theory
expanded around the universal solution, which has been found as an
analytic classical solution with one parameter, $a$. 
For $a>-1/2$, we are able to reproduce open string scattering amplitudes
in the theory fixed in the Siegel gauge.
At $a=-1/2$, all scattering amplitudes vanish and there is no open
string excitation in the gauge fixed theory.
These results support the conjecture that the universal
solution can be regarded as a pure gauge or the tachyon vacuum
solution. 
}

\begin{document}

\maketitle

\section{Introduction}

It has been conjectured that
the tachyon vacuum
solution\cite{rf:KS-tachyon,rf:SZ-tachyon,rf:MT,rf:GR} in open string
field theory has been 
analytically constructed as the universal solution that is found in the
universal subspace of string 
fields.\cite{rf:TT2}
This conjecture is strongly supported by the facts that
the modified BRS charge in the theory expanded around the universal
solution has vanishing cohomology in the Hilbert space of ghost number
one\cite{rf:KT} and that 
it has been shown numerically that the 
non-perturbative vacuum vanishes
in the theory.\cite{rf:tomo} These facts
imply that no open string excitation appears perturbatively
around the solution and that from the outset the theory is on the
tachyon vacuum.

There is an approach to describing the tachyon vacuum
other than finding the tachyon vacuum solution, namely  vacuum
string field theory.\cite{rf:RSZ-VSFT,rf:GRSZ} 
Though much progress has been
made in vacuum string field theory,\cite{rf:VSFT} a kind 
of regularization is needed due to the existence of a pure ghost
midpoint operator in 
the kinetic term.\cite{rf:GRSZ}
The kinetic operator of string field theory expanded around the
universal solution is less singular than this pure ghost
midpoint operator. Moreover, as Drukker has pointed out,\cite{rf:Drukker2}
the kinetic term constructed around the universal solution may become a
pure ghost midpoint operator in a certain singular limit.
It is thus evident that the theory expanded around the universal
solution is much more general and 
less singular than vacuum string field theory.

In order to prove the equivalence of the tachyon vacuum and the
universal solutions,
we must show that the energy density of the universal solution cancels
the D-brane tension exactly. In addition, we need to understand the
existence of closed strings in the theory expanded around the universal
solution. At present, it seems difficult to prove the exact cancellation
due to technical problems\cite{rf:tomo}, and therefore
it is necessary to adopt other approaches to understand the relation
between these solutions. To find closed strings, we should first
study gauge fixing and scattering amplitudes in the theory.

The universal solution $\Psi_0(a)$ given
in Ref.~\citen{rf:TT2} is believed to represent a pure gauge for
$a>-1/2$ and the tachyon vacuum solution at $a=-1/2$.
In Ref.~\citen{rf:Drukker2}, Drukker studies the
theory expanded around this solution fixed in the Siegel gauge and
suggests that 
original open string 
amplitudes can be reproduced around the solution for $a>-1/2$.
In this paper, 
we investigate this gauge-fixed theory more precisely and show more
explicitly the correspondence between the physical
states and amplitudes of the original and expanded theories.
Moreover, solving the cohomology under
the Siegel gauge condition, we prove that amplitudes become zero in
the theory expanded around the non-trivial solution at $a=-1/2$.

In \S 2,
we solve the equation of motion in the Siegel gauge
for $a>-1/2$ and we find
a one-to-one correspondence between the spectra of the original and
expanded theories. Next, we derive the
Feynman rule in the expanded theory. We find that this
Feynman rule yields the same scattering amplitudes as in the original
theory. In \S 3, we analyze the theory
expanded around the solution at $a=-1/2$ and we find that
the resulting scattering amplitudes become zero.
We give some discussion in \S 4.

\section{String field theory around pure gauge solutions}

\subsection{Pure gauge solutions}

The equation of motion in cubic open string field theory\cite{rf:CSFT}
is given by
\begin{eqnarray}
 \Q\Psi+\Psi*\Psi=0.
\end{eqnarray}
An analytic solution of the equation has been found in the
form\cite{rf:TT2} 
\begin{eqnarray}
\label{Eq:solution}
\Psi_0=Q_\rL(e^h-1)I-C_\rL((\partial h)^2 e^h) I,
\end{eqnarray}
where $I$ represents the identity string field, and the operators $Q_\rL$
and $C_\rL$ are defined as
\begin{eqnarray}
 Q_\rL(f)=\int_{C_{\rm left}} \frac{dw}{2\pi i}\,f(w)\,J_{\rm B}(w),
\ \ \ 
 C_\rL(f)=\int_{C_{\rm left}} \frac{dw}{2\pi i}\,f(w)\,c(w).
\end{eqnarray}
Here $J_{\rm B}(w)$ and $c(w)$ are the BRS current and the ghost field,
respectively. The function $h(w)$ in the solution satisfies
$h(-1/w)=h(w)$ and $h(\pm i)=0$. The solution can be expressed in terms
of the matter Virasoro generators and the ghost and anti-ghost oscillators
acting on the $SL(2,R)$ invariant vacuum, because it is constructed with
the BRS current, the ghost field and the identity string
field.\cite{rf:TT2} 
Because the expression of the solution does not depend on any specific
background, it is called the {\it universal solution}.

If we expand the string field $\Psi$ as
\begin{eqnarray}
 \Psi=\Psi_0+\Phi,
\end{eqnarray}
the action for the fluctuation $\Phi$ becomes
\begin{eqnarray}
 S[\Phi]=-\frac{1}{g^2}\int\left(
\frac{1}{2}\Phi*\Q'\Phi+\frac{1}{3}\Phi*\Phi*\Phi\right),
\end{eqnarray}
where the modified BRS charge is given by\cite{rf:TT2}
\begin{eqnarray}
\label{Eq:modBRS1}
 \Q'=Q(e^h)-C((\partial h)^2 e^h).
\end{eqnarray}
The operators $Q(f)$ and $C(f)$ are defined as
\begin{eqnarray}
 Q(f)=\oint \frac{dw}{2\pi i} f(w) J_{\rm B}(w),\ \ \ 
 C(f)=\oint \frac{dw}{2\pi i} f(w) c(w).
\end{eqnarray}

Let us consider the action expanded around the universal solution
generated by the function
\begin{eqnarray}
\label{Eq:ha}
h_a(w)=\log\left(1+\frac{a}{2}\left(w+\frac{1}{w}\right)^2\right),
\end{eqnarray}
where $a$ is a parameter larger than or equal to $-1/2$.\cite{rf:TT2}
Substituting this function into (\ref{Eq:modBRS1}),
we find that the modified BRS charge can be expanded as\cite{rf:TT2}
\begin{eqnarray}
\label{Eq:modBRS2}
\Q'(a)&=&(1+a)\Q+\frac{a}{2}(Q_2+Q_{-2})+4aZ(a)\,c_0
-2aZ(a)^2(c_2+c_{-2})\nn
&&-2a(1-Z(a)^2)
\sum_{n=2}^\infty (-1)^n Z(a)^{n-1}(c_{2n}+c_{-2n}),
\end{eqnarray}
where we have expanded the BRS current and the ghost field as $J_{\rm
B}(w)=\sum_n Q_n w^{-n-1}$ and $c(w)=\sum_n c_n w^{-n+1}$. The function
$Z(a)$ is defined by $Z(a)=(1+a-\sqrt{1+2a})/a$, and it varies
from $-1$ to $1$ for $a\geq -1/2$.

If the parameter $a$ is not equal to $-1/2$, the action can be
transformed into the action with the ordinary BRS charge through the string
field redefinition\cite{rf:TT2,rf:tomo}
\begin{eqnarray}
 \Phi'=e^{K(h_a)}\Phi,
\end{eqnarray}
where the operator $K(f)$ is defined by using the ghost number current
$J_{\rm gh}=cb$ as
\begin{eqnarray}
 K(f)=\oint\frac{dw}{2\pi i}f(w)\,
\left(J_{\rm gh}(w)-\frac{3}{2}\,w^{-1}\right).
\end{eqnarray}
Substituting the function (\ref{Eq:ha}) into this definition,
we obtain the mode expansion form of $K(h_a)$ as
\begin{eqnarray}
 K(h_a)=-\tilde{q}_0 \log(1-Z(a))^2
-\sum_{n=1}^\infty \frac{(-1)^n}{n}(q_{2n}+q_{-2n}) Z(a)^n,
\end{eqnarray}
where $\tilde{q}_0$ and $q_n$ are written in terms of the ghost
oscillators as 
\begin{eqnarray}
 \tilde{q}_0&=&\frac{1}{2}(c_0 b_0-b_0 c_0)
+\sum_{n=1}^\infty (c_{-n} b_n-b_{-n} c_n),\nn
 q_n &=& \sum_{m=-\infty}^\infty c_{n-m}b_m\ \ \ (n\neq 0).
\end{eqnarray}
Under the string field  redefinition, the modified BRS charge transforms
into the original one as follows:\cite{rf:TT2,rf:tomo}
\begin{eqnarray}
\label{Eq:simtrans}
\Q&\rightarrow&
\Q'=e^{K(h_a)}\,\Q(a)\,e^{-K(h_a)}.
\end{eqnarray}
Therefore we believe that the universal solution is pure gauge for
$a>-1/2$, as discussed in Ref.~\citen{rf:TT2}. Numerical analyses
strongly support this belief.\cite{rf:tomo}

Let us consider the BRS cohomology in
the theory expanded around the universal solution for $a>-1/2$.
As the cohomology of the Kato-Ogawa BRS charge, we know that any state
$\ket{\psi}$ satisfying $\Q\ket{\psi}=0$ can be written as 
\begin{eqnarray}
\label{Eq:cohomology1}
 \ket{\psi}=\ket{P}\otimes c_1 \ket{0}+
\ket{P'}\otimes c_0\,c_1 \ket{0}+\Q\ket{\phi},
\end{eqnarray}
where $\ket{P}$ and $\ket{P'}$ are positive norm states in the matter
sector,  and if we consider flat backgrounds
they are DDF states.\cite{rf:KatoOgawa,rf:Henneaux,rf:FGZ,rf:AsanoNatsu}
The perturbative equation of motion for the fluctuation is given by
$\Q'(a)\Phi=0$.
Because the modified and original BRS charges are related by the
similarity transformation (\ref{Eq:simtrans}), we can solve the equation
of motion to obtain the solution 
\begin{eqnarray}
\label{Eq:cohomology2}
 \ket{\Phi}=\ket{P}\otimes e^{K(h_a)}\,c_1 \ket{0}+
\ket{P'}\otimes e^{K(h_a)}\,c_0\,c_1 \ket{0}+\Q'(a)\ket{\phi}.
\end{eqnarray}
In this solution, the non-trivial cohomology parts possess the same
ghost number as the physical states in the original theory, and
$\ket{P}$ and $\ket{P'}$ in the matter sector do not change. 
This result suggests that the 
physical spectrum is the same as that in the original theory and 
it is natural for the universal solution with $a>-1/2$
to correspond to the pure gauge.

\subsection{Gauge fixing and the equation of motion}

Though we solved the cohomology above,
we must fix the gauge to determine the
physical spectrum precisely. Here, we apply the Siegel gauge
condition to the theory around the universal solution:
\begin{eqnarray}
 b_0\Phi=0.
\end{eqnarray}
Under the Siegel gauge condition, the equation of motion is given by
\begin{eqnarray}
\label{Eq:eqmotion}
L(a)\,\Phi=0,\ \ \ L(a)=\{\Q'(a),\,b_0\}.
\end{eqnarray}
From (\ref{Eq:modBRS2}), we can write the operator $L(a)$ in terms of an
oscillator expression as\cite{rf:tomo}
\begin{eqnarray}
\label{Eq:La1}
 L(a)=(1+a)L_0 +\frac{a}{2}(L_2+L_{-2})+a(q_2-q_{-2})+4aZ(a),
\end{eqnarray}
where the operators $L_n$ are the total Virasoro generators. We can
also rewrite $L(a)$ in terms of
the twisted ghost Virasoro generators,\cite{rf:GRSZ}
$L_n'=L_n+nq_n+\delta_{n,0}$, as
\begin{eqnarray}
\label{Eq:La2}
 L(a)=(1+a)L_0'+\frac{a}{2}(L_2'+L_{-2}')+4aZ(a)-1-a.
\end{eqnarray}
Because the twisted ghost conformal field theory has a central charge
$c'=24$, the three operators $L_0'+3$ and $L_{\pm2}'$ form an $SL(2,R)$
algebra. Then, it is useful to rewrite $L(a)$ as
\begin{eqnarray}
\label{Eq:La3}
 L(a)=2(1+a)l_0+a(l_2+l_{-2})+4aZ(a)-4(1+a),
\end{eqnarray}
where the operators $l_0$ and $l_{\pm 2}$ are defined by
\begin{eqnarray}
 l_0=\frac{1}{2}(L_0'+3),\ \ \ l_{\pm 2}=\frac{1}{2}L_{\pm 2}',
\end{eqnarray}
and these form the algebra defined by the relations
\begin{eqnarray}
\label{Eq:SLcomrel}
\left[l_0,\,l_{\pm 2}\right] = \mp l_{\pm 2},\ \ \ 
\left[l_2,\,l_{-2}\right] = 2l_0.
\end{eqnarray}

In order to solve the gauge fixed equation of motion
(\ref{Eq:eqmotion}), we 
attempt to diagonalize the operator $L(a)$. Because this operator
is expressed in terms of the $SL(2,R)$ generators as in
(\ref{Eq:La3}), it can be diagonalized under $SL(2,R)$
transformations. If we restrict the $SL(2,R)$ group to a subgroup
written by `normal ordered operators', arbitrary elements in the
subgroup can be represented as
\begin{eqnarray}
\label{Eq:SLelemnt}
 U(s,t,u)=\exp(s\,l_{-2})\,\exp(t\,l_0)\,\exp(u\,l_2),
\end{eqnarray}
where $s,t$ and $u$ are real parameters. Using the algebra
defined by (\ref{Eq:SLcomrel}), we can calculate the $SL(2,R)$
transformation of 
the operator $L(a)$ as
\begin{eqnarray}
 U(s,t,u)L(a)U(s,t,u)^{-1}
&=& \left\{2(1+a+au)-2s(au^2+2(1+a)u+a)\,e^{-t}\right\}\,l_0\nn
&&\hspace{-.5cm}
   +(au^2+2(1+a)u+a)\,e^{-t}\,l_2\nn
&&\hspace{-.5cm}
   +\left\{
    a\,e^t -2a su -2(1+a) s +s^2 (au^2+2(1+a)u+a)\,e^{-t}
    \right\}\,l_{-2}\nn
&&\hspace{-.5cm}
   +4aZ(a)-4(1+a).
\end{eqnarray}
If the coefficients of $l_{\pm 2}$ vanish, the
parameters $s,t$ and $u$ satisfy
\begin{eqnarray}
\label{Eq:stu1}
&& au^2+2(1+a)u+a=0, \\
\label{Eq:stu2}
&& a\,e^t-2a su -2(1+a) s =0.
\end{eqnarray}
From (\ref{Eq:stu1}), we find that either $u=-Z(a)$ or
$u=-1/Z(a)$. Substituting 
these values of $u$ into (\ref{Eq:stu2}), we obtain
\begin{eqnarray}
\label{Eq:stu3}
 a\,e^t \mp 2\sqrt{1+2a}\,s=0.
\end{eqnarray}
It should be noted that there is no solution of the
equation (\ref{Eq:stu3}) if $a=-1/2$ and $t$ is a finite real
number. This fact implies that the operator $L(a)$ cannot be transformed
into a form linear in  $l_0$ under
regular $SL(2,R)$ transformations. This difference between the cases
$a>-1/2$ and $a=-1/2$ is natural, because the theory
should have different physical 
spectrum in each case if our conjecture regarding the universal solution
holds.

In the case $a>-1/2$, we can set $s=-u$ in order to
diagonalize  $L(a)$. 
In this case, the operator $L(a)$ can be transformed into
$L_0$ as
\begin{eqnarray}
\label{Eq:Ldiag}
 U'(a)L(a)U'(a)^{-1}=\sqrt{1+2a}\,L_0,
\end{eqnarray}
where $U'(a)$ is given by
\begin{eqnarray}
\label{Eq:Ua}
 U'(a)&=&\exp\left\{\frac{1}{2}\,Z(a)L_{-2}'\right\}
      \exp\left\{\frac{1}{2}(L_0'+3)\log(1-Z(a)^2)\right\}
      \exp\left\{-\frac{1}{2}\,Z(a)L_2'\right\}\nn
     &=&\exp\left\{
     -\frac{1}{4}\left(L_2'-L_{-2}'\right)
    \log\left(\frac{1+Z(a)}{1-Z(a)}\right)\right\}.
\end{eqnarray}
Therefore, we obtain th solution
$\Phi={U'(a)}^{-1} \Phi_0$ of the gauge fixed equation of motion
(\ref{Eq:eqmotion}),
where $\Phi_0$ satisfies $L_0 \Phi_0=0$.

\begin{figure}
\centerline{\includegraphics[width=14cm]{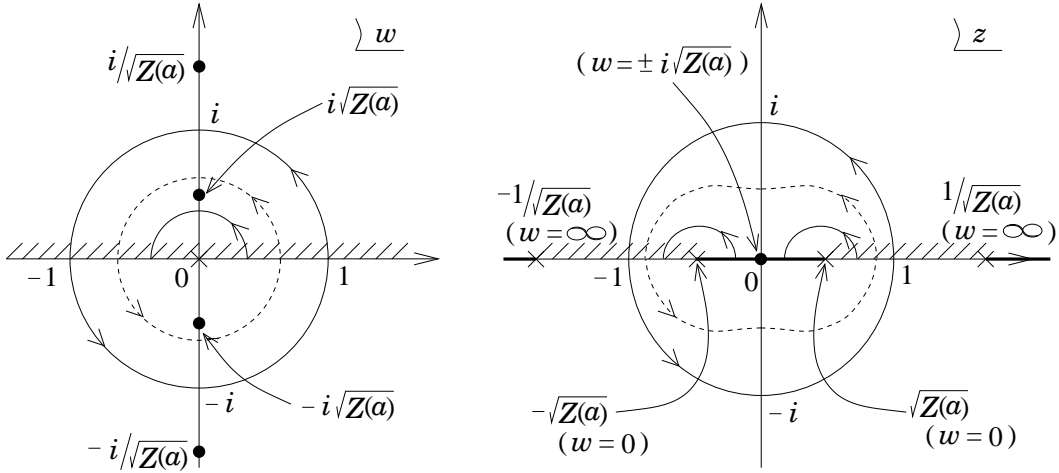}}
\caption{Conformal mapping of the $w$ plane into the $z$ plane under the
 mapping $z=f_a(w)$ in the case $0<Z(a)<1\ (a>0)$.}
\label{fig:map1}
\end{figure}

For later convenience, we present here another procedure for the
diagonalization of $L(a)$ making use of conformal field theory.
In the twisted ghost conformal field theory, the operator $U'(a)$ induces
a conformal mapping represented by the function\cite{rf:LPP1}
\begin{eqnarray}
\label{Eq:fa}
f_a(w)=\left(\frac{w^2+Z(a)}{Z(a)\,w^2+1}\right)^{\frac{1}{2}};
\end{eqnarray}
that is, if $\phi(w)$ is a primary field of dimension $h$, the
transformation of $\phi(w)$ by $U'(a)$ is given by
\begin{eqnarray}
 U'(a)\phi(w)U'(a)^{-1} = \left(\frac{df_a(w)}{dw}\right)^h
\phi\left(f_a(w)\right).
\end{eqnarray}
For $0<Z(a)<1\ (a>0)$, this conformal mapping is depicted in
Fig.~\ref{fig:map1}. In the $z$ plane, there is a branch cut connecting
$-\sqrt{Z(a)}$ and $\sqrt{Z(a)}$. This branch cut corresponds to a line
segment on the imaginary axis between $-i\sqrt{Z(a)}$ and
$i \sqrt{Z(a)}$ in the $w$ plane. There is another branch cut from $\pm
1/\sqrt{Z(a)}$ to infinity in the $z$ plane. 
For $-1<Z(a)<0\ (-1/2<a<0)$, the mapping is represented by the figure
that is obtained by rotating Fig.~\ref{fig:map1} clockwise $90^\circ$.

From (\ref{Eq:La2}), we find that the operator $L(a)$ can be expressed
in terms of the 
twisted energy momentum tensor $T'(w)=\sum L_n' w^{-n-2}$ as
\begin{eqnarray}
\label{Eq:LT}
 L(a)=\oint \frac{dw}{2\pi i}\,w\,e^{h_a(w)}\,T'(w)
+4Z(a)-1-a,
\end{eqnarray}
where the integration contour is a unit circle.
Generally, the kinetic operator can be written using the twisted
energy momentum tensor in the theory around universal
solutions in the Siegel gauge.\cite{rf:Drukker2} 
Then, the transformation of $L(a)$ by $U'(a)$ is given by
\begin{eqnarray}
\label{Eq:ULTU}
 U'(a)L(a)U'(a)^{-1} 
=
\oint \frac{dw}{2\pi i}\,w\,e^{h_a(w)}\,U'(a)T'(w)U'(a)^{-1}
+4Z(a)-1-a.
\end{eqnarray}
Under the conformal mapping $z=f_a(w)$,
the twisted energy momentum tensor is transformed to
\begin{eqnarray}
\label{Eq:UTU}
 U'(a)T'(w)U'(a)^{-1}=
 \left(\frac{df_a(w)}{dw}\right)^2 T'(f_a(w))+\frac{c'}{12} S(f_a,w),
\end{eqnarray}
where the central charge $c'$ is $24$ and $S(f,w)$ denotes the
Schwartzian derivative:
\begin{eqnarray}
 S(f,w)=\frac{\partial^3 f(w)}{\partial f(w)}-\frac{3}{2}\left(
\frac{\partial^2 f(w)}{\partial f(w)}\right)^2.
\end{eqnarray}
Using (\ref{Eq:ha}) and (\ref{Eq:fa}), we see that
\begin{eqnarray}
&&
 e^{h_a(w)}=\frac{1}{(1-Z(a))^2}\,w^{-2}\,(w^2+Z(a))(Z(a)w^2+1), \\
&&
 S(f_a,w)=-\frac{3Z(a)}{2}\,w^{-2}\,
\frac{(1+2Z(a)w^2+w^4)(Z(a)+2w^2+Z(a)w^4)}{
(w^2+Z(a))^2(Z(a)w^2+a)^2},
\end{eqnarray}
and therefore
\begin{eqnarray}
\label{Eq:ehf}
&&
 w\,e^{h_a(w)}\,f_a'(w)=\frac{1+Z(a)}{1-Z(a)}\,f_a(w), \\
\label{Eq:ehS}
&&
 w\,e^{h_a(w)}\,S(f_a,w)
=-\frac{3Z(a)}{2(1-Z(a))^2}\nn
&&\hspace{4cm}\times
\frac{(1+2Z(a)w^2+w^4)(Z(a)+2w^2+Z(a)w^4)}{w^3\,
(w^2+Z(a))(Z(a)w^2+a)}.
\end{eqnarray}
Substituting (\ref{Eq:ULTU}) into (\ref{Eq:UTU}) and then using
(\ref{Eq:ehf}) and (\ref{Eq:ehS}), we find
\begin{eqnarray}
\label{Eq:ULU2}
 U'(a)L(a)U'(a)^{-1}
&=&
\frac{1+Z(a)}{1-Z(a)}\,
\oint \frac{dw}{2\pi i}\,f_a(w)\,\frac{df_a(w)}{dw}\,T'(f_a(w)) \nn
&&
-\frac{3Z(a)}{(1-Z(a))^2}
\oint \frac{dw}{2\pi i}\,
\frac{(1+2Z(a)w^2+w^4)(Z(a)+2w^2+Z(a)w^4)}{w^3\,
(w^2+Z(a))(Z(a)w^2+a)} 
\nn
&&
+4Z(a)-1-a.
\end{eqnarray}
In the first term on the right-hand side of this expression, we can
change the variable of 
integration to $z=f_a(w)$, since the integration contour
does not cross the branch cut in the $z$ plane, as seen in
Fig.~\ref{fig:map1}.  Then, the first term becomes
\begin{eqnarray}
\label{Eq:term1}
 \frac{1+Z(a)}{1-Z(a)}\,
\oint \frac{dz}{2\pi i}\,z\,T'(z)
=\sqrt{1+2a}\,L_0'. 
\end{eqnarray}
If $1>Z(a)>0$, the integral in the second term in (\ref{Eq:ULU2}) can be
reduced to the 
summation of the residues at $w=0$ and $\pm i \sqrt{Z(a)}$,
which individually are
\begin{eqnarray}
\label{Eq:term2}
&&
\oint_{C_0} = \frac{1+Z(a)^2}{Z(a)},
\ \ \ 
\oint_{C_{i\,\sqrt{Z(a)}}}=
\oint_{C_{-i\,\sqrt{Z(a)}}}
=\frac{-1+Z(a)^2}{2 Z(a)}.
\end{eqnarray}
If $-1<Z(a)<0$, the residues at $\pm i \sqrt{Z(a)}$ are replaced by
those at $\pm \sqrt{-Z(a)}$, but their values are unchanged.
Finally, substituting (\ref{Eq:term1}) and (\ref{Eq:term2}) into
(\ref{Eq:ULTU}), we can derive the same result as obtained from operator
expressions:
\begin{eqnarray}
 U'(a)L(a)U'(a)^{-1}
&=&\sqrt{1+2a}\,L_0'-\frac{3 Z(a)}{(1-Z(a))^2}\,2Z(a)
+4Z(a)-1-a \nn
&=& \sqrt{1+2a}\,L_0.
\end{eqnarray}
Note that this result can not be applied to the case $a=-1/2$,
because the operator $U'(a)$ becomes singular at $a=-1/2$ as seen in
(\ref{Eq:Ua}).

\subsection{Physical subspace}

First, we show that the modified BRS charge is transformed
into the original one through the similarity transformation with $U'(a)$:
\begin{eqnarray}
\label{Eq:UQU1}
 U'(a)\Q'(a)U'(a)^{-1}=\sqrt{1+2a}\,\Q.
\end{eqnarray}

The commutation relations of $L_m$ and $q_m$ with $Q_n$ are given by
\begin{eqnarray}
&& \left[L_m,\,Q_n\right]=-nQ_{m+n},\nn
&&
\left[q_m,\,Q_n\right]=Q_{m+n}-2m n\,c_{m+n}.
\end{eqnarray}
Then, we obtain the commutation relation of $L_m'$ with $Q_n$ as
\begin{eqnarray}
 \left[L_m',\,Q_n\right]=
(m-n)Q_{m+n}-2m^2 n\,c_{m+n}.
\end{eqnarray}
The first term on the left-hand side implies that $Q_n$ is transformed as
oscillators of 
a dimension two field in the twisted ghost theory.
Due to the second term, $Q_n$
is not an oscillator of a primary field.
This is a natural result of the fact that
the BRS current can be written $J_{\rm B}(w)=w c'(w) T_X(w)+\cdots$,
where $c'(w)$ and $T_X(w)$ denote the ghost field of dimension zero and
the matter energy momentum tensor of dimension two in the twisted
theory, and the dots stand for terms containing no matter
oscillators.

From the above, we see that, because $\Q'(a)$ and $L(a)$ have similar
oscillator expressions,
\begin{eqnarray}
 \Q'(a)&=&(1+a)\Q+\frac{a}{2}(Q_2+Q_{-2})+\cdots,\\
 L(a)&=&(1+a)L_0'+\frac{a}{2}(L_2'+L_{-2}')+\cdots,
\end{eqnarray} 
and the operators $L_n'$ are oscillators of a dimension 2 field, we can
obtain the 
transformation of $\Q'(a)$ by analogy to the previous
result for $L(a)$ as
\begin{eqnarray}
\label{Eq:UQU2}
 U'(a)\,\Q'(a)U'(a)^{-1}
= \sqrt{1+2a}\,\Q +\cdots,
\end{eqnarray}
where the dots represent pure ghost contributions.
Because the right-hand side of (\ref{Eq:UQU2}) is nilpotent, the pure
ghost contributions are found to be zero. This follows from the fact
that an operator of the form $Q(f)+C(g)$ must be $Q(e^f)-C((\partial
f)^2 e^f)$ or $C(f)$ if it 
is nilpotent.\cite{rf:KT} Therefore, we have shown that the
transformation law given in (\ref{Eq:UQU1}) holds.

Here, we should discuss the cohomology without the Siegel gauge
condition. Using the equation (\ref{Eq:UQU1}), we show that the
state $\ket{\Phi}$ with the condition $\Q'(a)\ket{\Phi}=0$ can be
written as
\begin{eqnarray}
\label{Eq:cohomology4}
 \ket{\Phi}=U'(a)^{-1}
(\ket{P}\otimes \,c_1 \ket{0})+
U'(a)^{-1}
(\ket{P'}\otimes \,c_0\,c_1 \ket{0})+\Q'(a)\ket{\phi}.
\end{eqnarray}
However, this expression is different from that for the state
given in (\ref{Eq:cohomology2}), though these states are imposed by the
same condition. We can show with a natural line of reasoning that the
apparent difference between
these states reduces to merely a BRS-exact state.

In order to see this fact, we prove that
\begin{eqnarray}
\label{Eq:UfUfK}
 U'_f={\rm const}\times U_f\times \exp K\left(
\log \frac{w\,\partial f(w)}{f(w)}\right),
\end{eqnarray}
where $U'_f$ and $U_f$ are the operators that implement the conformal
transformation $z=f(w)$ in the twisted and untwisted ghost conformal
field theory, respectively.

In the untwisted theory, the ghost and anti-ghost fields
are expanded in a unit disc as 
\begin{eqnarray}
 c(w)=\sum_{n=-\infty}^\infty c_n w^{-n+1},\ \ \ 
 b(w)=\sum_{n=-\infty}^\infty b_n w^{-n-2},
\end{eqnarray}
and in the twisted theory they are expanded as
\begin{eqnarray}
  c'(w)=\sum_{n=-\infty}^\infty c_n w^{-n},\ \ \ 
 b'(w)=\sum_{n=-\infty}^\infty b_n w^{-n-1}.
\end{eqnarray}
Then, the ghost and anti-ghost fields in both theories can be
connected by the relations 
\begin{eqnarray}
\label{Eq:bcrel}
 c'(w)=w^{-1}\,c(w),\ \ \ b'(w)=w\,b(w).
\end{eqnarray}
For these ghost fields, the operators $U_f$ and $U'_f$ induce the
transformations
\begin{eqnarray}
\label{Eq:UcU1}
 U_f\,c(w)\,U_f^{-1} &=&
   \left(\partial f(w)\right)^{-1} c\left(f(w)\right),\\
\label{Eq:UbU1}
 U_f\,b(w)\,U_f^{-1} &=&
   \left(\partial f(w)\right)^2 b\left(f(w)\right),\\
\label{Eq:UcU2}
 U'_f\,c'(w)\,{U'_f}^{-1} &=&
   \left(\partial f(w)\right)^0 c'\left(f(w)\right),\\
\label{Eq:UbU2}
 U'_f\,b'(w)\,{U'_f}^{-1} &=&
   \left(\partial f(w)\right)^1 b'\left(f(w)\right).
\end{eqnarray}
Combining (\ref{Eq:bcrel}) and (\ref{Eq:UcU1}), we can
find the transformation of $c'(w)$ through $U_f$:
\begin{eqnarray}
\label{Eq:UcU3}
 U_f\,c'(w)\,U_f^{-1} &=& 
 w^{-1}\,U_f\,c(w)\,U_f^{-1}\nn
&=&
 w^{-1}\,(\partial f(w))^{-1}\,c(f(w))\nn
&=&
 \frac{f(w)}{w\partial f(w)}\,c'(f(w)).
\end{eqnarray}
Similarly, from (\ref{Eq:bcrel}) and (\ref{Eq:UcU1}), the transformation
of $b(w)$ is given by 
\begin{eqnarray}
\label{Eq:UbU3}
 U_f\,b'(w)\,{U_f}^{-1} &=& \frac{w(\partial f(w))^2}{f(w)}
\,b'(f(w)).
\end{eqnarray}
We can realize the same transformations of (\ref{Eq:UcU3}) and
(\ref{Eq:UbU3}) using $U_f\,e^{K(g)}$.
For an arbitrary function $g(w)$, the operator $K(g)$ generates
the transformations\cite{rf:TT2,rf:Drukker2,rf:tomo}
\begin{eqnarray}
\label{Eq:KcK}
 e^{K(g)}\,c'(w)\,e^{-K(g)} &=& e^{g(w)}\,c'(w),\\
\label{Eq:KbK}
 e^{K(g)}\,b'(w)\,e^{-K(g)} &=& e^{-g(w)}\,b'(w).
\end{eqnarray}
Combining (\ref{Eq:UcU3}), (\ref{Eq:UbU3}), (\ref{Eq:KcK}) and
(\ref{Eq:KbK}), we find
\begin{eqnarray}
\label{Eq:UKcKU}
 U_f e^{K(g)}\,c'(w)\,e^{-K(g)} U_f^{-1} &=&
 e^{g(w)}\,\frac{f(w)}{w(\partial f(w))}\,c'(f(w)), \\
\label{Eq:UKbKU}
 U_f e^{K(g)}\,b'(w)\,e^{-K(g)} U_f^{-1} &=&
 e^{-g(w)}\,\frac{w(\partial f(w))^2}{f(w)} b'(f(w)).
\end{eqnarray}
Comparing (\ref{Eq:UcU2}) and (\ref{Eq:UbU2}) with (\ref{Eq:UKcKU}) and
(\ref{Eq:UKbKU}), we find that these transformation laws are equivalent
if $g(w)$ is given by
\begin{eqnarray}
 g(w)=\log\left(\frac{w\,\partial f(w)}{f(w)}\right).
\end{eqnarray}
The ghost part of $U'_f$
is uniquely determined by the transformation laws
(\ref{Eq:UcU2}) and (\ref{Eq:UbU2}), up to a multiplicative constant.
The matter parts of $U_f$ and $U'_f$ have a same form, because they are
uncharged by the twist operation. 
Therefore the equation (\ref{Eq:UfUfK}) is proved.
In Appendix A, we generalize this equation to the case of general
background charge.

We now apply the formula (\ref{Eq:UfUfK}) to the mapping $f(w)=f_a(w)$
given by (\ref{Eq:fa}). We can calculate the function in the operator
$K$ as
\begin{eqnarray}
 \log\left(
\frac{w \partial f_a(w)}{f_a(w)}
\right) &=&
\log\left\{
   (1-Z(a)^2) \frac{w^2}{(w^2+Z(a))(Z(a)w^2+1)}
\right\} \nn
&=& \log\left\{
   (1-Z(a))^2 \frac{w^2}{(w^2+Z(a))(Z(a)w^2+1)}
\right\}+\log\frac{1+Z(a)}{1-Z(a)}
\nn
&=& -h_a(w)+\log\frac{1+Z(a)}{1-Z(a)}.
\end{eqnarray}
Therefore, we can represent the operator $U'(a)$ in terms of the
untwisted Virasoro generators as
\begin{eqnarray}
\label{Eq:UaUaK}
 U'(a)= U(a)\,e^{-K(h_a)}\,\exp\left(\tilde{q}_0
\log\frac{1+Z(a)}{1-Z(a)}\right),
\end{eqnarray}
where the operator $U(a)$ is defined as
\begin{eqnarray}
\label{Eq:Ua2}
 U(a) &=&\exp\left\{
     -\frac{1}{4}\left(L_2-L_{-2}\right)
    \log\left(\frac{1+Z(a)}{1-Z(a)}\right)\right\} \nn
&=&\exp\left\{\frac{1}{2}\,Z(a)L_{-2}\right\}
      \exp\left\{\frac{1}{2}\,L_0\log(1-Z(a)^2)\right\}
      \exp\left\{-\frac{1}{2}\,Z(a)L_2\right\}.
\end{eqnarray}
Here, the multiplicative constant is 1 because
$U'(a)$ is a unitary operator.\footnote{The hermiticity property
is given by $(q_n)^\dagger=-q_{-n}\ (n\neq 0)$,
$(\tilde{q}_0)^\dagger=-\tilde{q}_0$ and $(L_n)^\dagger=L_{-n}$.}
Since the untwisted Virasoro generators commute with the BRS charge, it
is easily seen from (\ref{Eq:UaUaK}) that (\ref{Eq:simtrans}) and
(\ref{Eq:UQU1}) hold simultaneously.

Using (\ref{Eq:UaUaK}), we can express the operator $U'(a)^{-1}$ as
\begin{eqnarray}
 U'(a)^{-1} &=&
 e^{K(h_a)}\,U(a)^{-1}\exp\left(-\tilde{q}_0
\log\frac{1+Z(a)}{1-Z(a)}\right)
\nn
&=&
 e^{K(h_a)}\times \sum_{n=0}^\infty
\frac{1}{n!}(L_2-L_{-2})^n \theta^n \times
e^{-4 \tilde{q}_0\,\theta}
\ \ \ \left(
\theta=
\frac{1}{4}\log\frac{1+Z(a)}{1-Z(a)}
\right)
\nn
&=&
 e^{K(h_a)}\,e^{-4 \tilde{q}_0 \theta} \nn
&& + e^{K(h_a)}\times
\left\{\Q,\,
\sum_{n=1}^\infty
\frac{1}{n!}(b_2-b_{-2})(L_2-L_{-2})^{n-1} \theta^n 
\right\}
\times
e^{-4 \tilde{q}_0 \theta}
\end{eqnarray}
Then we can rewrite the first term of (\ref{Eq:cohomology4}) as
\begin{eqnarray}
\label{Eq:UPKP1}
 U'(a)^{-1}(\ket{P}\otimes c_1 \ket{0}) &=&
e^{2\theta}\ket{P}\otimes e^{K(h_a)} c_1\ket{0}\nn
&&
+e^{K(h_a)}\Q
\sum_{n=1}^\infty
\frac{1}{n!}(b_2-b_{-2})(L_2-L_{-2})^{n-1} \theta^n 
(e^{2\theta}\ket{P}\otimes c_1\ket{0}) \nn
&=&
e^{2\theta}
\ket{P}\otimes e^{K(h_a)} c_1\ket{0}
+\Q'(a)\ket{\phi},
\end{eqnarray}
where $\ket{\phi}$ is given by
\begin{eqnarray}
\ket{\phi}= e^{K(h_a)}\sum_{n=1}^\infty
\frac{1}{n!}(b_2-b_{-2})(L_2-L_{-2})^{n-1} \theta^n 
(e^{2\theta}\ket{P}\otimes c_1\ket{0}).
\end{eqnarray}
Similarly, the second term of (\ref{Eq:cohomology4}) can be
written
\begin{eqnarray}
\label{Eq:UPKP2}
 U'(a)^{-1}(\ket{P'}\otimes c_0 c_1\ket{0})
&=& 
e^{-2\theta}
\ket{P'}\otimes e^{K(h_a)} c_0 c_1\ket{0}
+\Q'(a)\ket{\phi'}.
\end{eqnarray}
Hence, from (\ref{Eq:UPKP1}) and (\ref{Eq:UPKP2}),
it is shown that the difference between the cohomologies
(\ref{Eq:cohomology2}) and (\ref{Eq:cohomology4})
can be expressed as a BRS-exact state, as asserted above.

Now that the cohomology of $\Q'(a)$ has been established, we can
obtain the physical subspace specified by the conditions
\begin{eqnarray}
\label{Eq:Qbphi}
 \Q'(a)\ket{\Phi}=0,\ \ \ b_0\ket{\Phi}=0.
\end{eqnarray}
Because, under the same similarity transformation,  $\Q'(a)$ and $L(a)$
are transformed into $\Q$ and $L_0$, the state $\ket{\Phi}$
satisfying (\ref{Eq:Qbphi}) can be written
\begin{eqnarray}
\label{Eq:cohomology3}
 \ket{\Phi}=U'(a)^{-1}\left(
\ket{P}\otimes c_1\ket{0}
+ \tQ \ket{\phi} \right),
\end{eqnarray}
where $\tQ$ denotes the terms of $\Q$ that do not contain the ghost and
anti-ghost zero modes, $c_0$ and $b_0$.
Here, we have used the cohomology for the Kato-Ogawa BRS charge
(\ref{Eq:cohomology1}) and the commutation relations $[L_n',\,b_0]=0$.
Hence, there is a one-to-one correspondence between the spectra of the
original theory and those of the
theory around the universal solution for $a>-1/2$. They are connected
through the similarity 
transformation with $U'(a)$.

\subsection{Scattering amplitudes}

Here, we consider scattering amplitudes in the theory around
the universal solution. 
The general amplitude is calculated by evaluating an expression of
the form
\begin{eqnarray}
 {\cal A} = 
\left(\prod \bra{V}\right)\left(\prod \frac{b_0}{L(a)}\right)
\left(\prod \ket{R}\right)
 \left(\prod\ket{\rm external}' \right),
\end{eqnarray}
where $\bra{V}$, $\ket{R}$ and $\ket{\rm external}'$ are vertices,
reflectors and external states, respectively. Also, $b_0/L(a)$ is
the propagator in the theory expanded around the universal
solution. From (\ref{Eq:Ldiag}) 
and $[L_n',\,b_0]=0$, we 
can rewrite the propagator as
\begin{eqnarray}
\label{Eq:prop}
 \frac{b_0}{L(a)} 
&=& \frac{1}{\sqrt{1+2a}}\times {U'(a)}^{-1}\,\frac{b_0}{L_0}\,
U'(a).
\end{eqnarray}
From (\ref{Eq:cohomology3}), the external states can be
written as  
similarity transformations of the external states in the original
theory: 
\begin{eqnarray}
\label{Eq:extstate}
\ket{\rm external}'={U'(a)}^{-1}\ket{\rm external}.
\end{eqnarray}
It can be easily seen that on the reflector, the operator
$K'_n=L'_n-(-1)^n L'_{-n}$
satisfies
\begin{eqnarray}
 {}_{12}\bra{R}({K'_n}^{(1)}+{K'_n}^{(2)})=0,
\end{eqnarray}
and then we find
\begin{eqnarray}
\label{Eq:reflector}
{}_{12}\bra{R}\prod_{r=1}^2\,(U'(a)^{(r)})^{-1}
={}_{12}\bra{R}.
\end{eqnarray}
Using (\ref{Eq:prop}), (\ref{Eq:extstate}) and (\ref{Eq:reflector}), 
we can rewrite the amplitude into the form
\begin{eqnarray}
 {\cal A} = 
\left(\prod \bra{V'} \right)\left(\prod \frac{1}{\sqrt{1+2a}} 
\frac{b_0}{L_0}\right)
\left(\prod \ket{R}\right)
 \left(\prod\ket{\rm external} \right).
\end{eqnarray}
The difference between this and the original amplitude is in the
normalization factor 
of the propagator and the change of the vertex
${}_{123}\bra{V}$ to
\begin{eqnarray}
{}_{123}\bra{V'} ={}_{123}\bra{V}\prod_{r=1}^3{U'(a)^{(r)}}^{-1}.
\end{eqnarray}

The operator $K_n=L_n-(-1)^n L_{-n}$ is conserved on the
original vertex.\cite{rf:SSFT}
In addition,  we find that the operator $K(h_a)$ also is conserved on
the original vertex.\cite{rf:TT2} Using these conservation laws and the
expression  
(\ref{Eq:UaUaK}) of $U'(a)$,
we find that the
modified vertex can be rewritten as
\begin{eqnarray}
 {}_{123}\bra{V'}=
{}_{123}\bra{V}\prod_{r=1}^3\,
\exp\left(-\log \sqrt{1+2a}\,{\tilde{q}_0}^{(r)}\right).
\end{eqnarray}
Under the $SL(2,R)$ normal ordering, the zero mode $q_0$ of the ghost
number current yields an anomalous contribution on the vertex:\cite{rf:RZ}
\begin{eqnarray}
 {}_{123}\bra{V}\sum_{r=1}^3{q_0}^{(r)}=
 3\times{}_{123}\bra{V}.
\end{eqnarray}
Since $\tilde{q}_0=q_0-3/2$,  we find
\begin{eqnarray}
{}_{123} \bra{V'}=\left(
\sqrt{1+2a}\right)^{\textstyle \frac{3}{2}}\times{}_{123}\bra{V}.
\end{eqnarray}

Combining the above results, we find that the perturbative amplitude in
the theory around the solution takes the form
\begin{eqnarray}
 {\cal A} = 
\left(\prod \left(
\sqrt{1+2a}\right)^{\textstyle \frac{3}{2}}
\bra{V} \right)\left(\prod \frac{1}{\sqrt{1+2a}} 
\frac{b_0}{L_0}\right)
\left(\prod \ket{R}\right)
 \left(\prod\ket{\rm external} \right). 
\end{eqnarray}
The normalization factors of the propagators and the vertices
cancel and the remaining factors can be
absorbed into the normalization of the external states as
$(\sqrt{1+2a})^{1/2}\times \ket{\rm external}$. 
Finally, it is found that the amplitude becomes equal to the
corresponding amplitude 
in the original theory. 
Hence, we conclude that the theory for $a>-1/2$
describes the same physics as the theory with the original BRS charge.

\section{String field theory around non-trivial solutions}

\subsection{Non-trivial solutions}

In this section we consider the theory expanded around the universal
solution which is obtained by setting the parameter $a$ to $-1/2$.
At $a=-1/2$, the function (\ref{Eq:ha}) is given by
\begin{eqnarray}
 h(w)=\log\left(-\frac{1}{4}\left(w-\frac{1}{w}\right)^2\right),
\end{eqnarray}
and the modified BRS charge (\ref{Eq:modBRS2}) becomes
\begin{eqnarray}
 \bQ &\equiv& \Q'(-1/2) \nn
   &=&
\frac{1}{2}\Q-\frac{1}{4}(Q_2+Q_{-2})
+2c_0 +c_2+c_{-2}.
\end{eqnarray}
Through a similarity transformation, this modified BRS charge can be
transformed into a form that contains
specific level oscillators,\cite{rf:KT}
\begin{eqnarray}
\label{Eq:KbQK}
 e^{K(\rho)}\bQ e^{-K(\rho)}=
-\frac{1}{4}Q_2+c_2,
\end{eqnarray}
where $\rho(w)$ and $K(\rho)$ are given by
\begin{eqnarray}
\label{Eq:rho}
&&
 \rho(w)=-2\log(1-w^{-2})
=2\sum_{n=1}^\infty \frac{1}{n}w^{-2n}, \\
\label{Eq:Krho}
&&
 K(\rho)=2\sum_{n=1}^\infty \frac{1}{n}q_{-2n}.
\end{eqnarray}
Here we introduce an operator $O^{(k)}$ to a certain operator
$O$ that is defined by replacing the ghost oscillator modes $c_n$
and $b_n$ in $O$ by $c_n^{(k)}=c_{n+k}$ and $b_n^{(k)}=b_{n-k}$
without changing their order.\cite{rf:KT}\footnote{
Note that $O^{(k)}$ here does {\it not} represent the operator $O$ of
the {\it k}-th string.}
With this definition, we can rewrite the equation (\ref{Eq:KbQK}) as
\begin{eqnarray}
\label{Eq:KbQK2}
\bQ =
-\frac{1}{4}\,e^{-K(\rho)}\,\Q^{(2)}\,e^{K(\rho)}.
\end{eqnarray}
For this $bc$-shift transformation, it is important that
the original algebra of the operator $O$ is realized to
the operator $O^{(k)}$. This follows from
the anti-commutation relations
$\{c_m^{(k)},\,b_n^{(k)}\}=\delta_{m+n,0}$. Consequently, through the
relation (\ref{Eq:KbQK2}), we can determine 
the cohomology of $\bQ$ by referring the original cohomology of $\Q$. We
find that any state $\ket{\psi}$ satisfying $\bQ\ket{\psi}=0$ can be
written\cite{rf:KT}
\begin{eqnarray}
\label{Eq:cohom1}
 \ket{\psi}=\ket{P}\otimes e^{-K(\rho)}b_{-2}\ket{0}
+\ket{P'}\otimes e^{-K(\rho)}\ket{0}+\bQ\ket{\phi}.
\end{eqnarray}
Here, the states $b_{-2}\ket{0}$ and $\ket{0}$ correspond to 
$c_1\ket{0}$ and $c_0 c_1\ket{0}$ in the original cohomology
(\ref{Eq:cohomology1}), respectively.

In gauge unfixed string field theory, all component fields of the string
field correspond to states of ghost number one. Therefore, the
resulting cohomology (\ref{Eq:cohom1}) implies that all on-shell
modes are reduced to gauge degrees of freedom in the gauge unfixed
theory.\cite{rf:KT}

\subsection{No open string theorem}

First, we demonstrate the formula
\begin{eqnarray}
\label{Eq:UfUfKk}
 U_f^{(k)}={\rm const}\times U'_f\times
\exp K\left(\log\,\frac{f(w)^{k+1}}{w^{k+1} \partial f(w)}\right).
\end{eqnarray}
Because the $bc$-shift transformation preserves the forms of commutation
relations, the operators $c^{(k)}(w)$ and $b^{(k)}(w)$ are transformed
as primary fields of dimension $-1$ and $2$ under the similarity
transformation with $U_f^{(k)}$:
\begin{eqnarray}
\label{Eq:UcUk}
&& U_f^{(k)}\,c^{(k)}(w)\,{U_f^{(k)}}^{-1}
=(\partial f(w))^{-1}\,c^{(k)}(w),\\ 
\label{Eq:UbUk}
&& U_f^{(k)}\,b^{(k)}(w)\,{U_f^{(k)}}^{-1}
=(\partial f(w))^{2}\,b^{(k)}(w).
\end{eqnarray}
The operators $c^{(k)}(w)$ and $b^{(k)}(w)$ can be written in terms of
the twisted operators as
\begin{eqnarray}
\label{Eq:cck}
&&
 c^{(k)}(w)=\sum_{n=-\infty}^\infty c_{n+k} w^{-n+1}
           =w^{k+1}\,c'(w),\\
\label{Eq:bbk}
&&
 b^{(k)}(w)=\sum_{n=-\infty}^\infty b_{n-k} w^{-n-2}
           =w^{-k-1}\,b'(w).
\end{eqnarray}
Then, from these relations and the equations (\ref{Eq:UcU1}),
(\ref{Eq:UbU1}), (\ref{Eq:KcK}) and (\ref{Eq:KbK}), it follows that
\begin{eqnarray}
 U'_f\,e^{K(g)}\,c^{(k)}(w)\,e^{-K(g)}\,{U'_f}^{-1}
&=&  w^{k+1} e^{g(w)} (f(w))^{-k-1} c^{(k)}(f(w)), \\
 U'_f\,e^{K(g)}\,b^{(k)}(w)\,e^{-K(g)}\,{U'_f}^{-1}
&=& w^{-k-1} e^{-g(w)}\,\partial f(w)\, (f(w))^{k+1} \,b^{(k)}(f(w)).
\end{eqnarray}
If the function $g(w)$ is given by
\begin{eqnarray}
\label{Eq:gk}
 g(w)=\log\,\frac{f(w)^{k+1}}{w^{k+1} \partial f(w)},
\end{eqnarray}
these transformation laws coincide with the similarity transformation with
$U_f^{(k)}$. Because the $bc$-shift transformation does not affect
matter oscillators, the operators $U_f^{(k)}$ and $U_f e^{K(g)}$, in which
$g(w)$ is given by (\ref{Eq:gk}), are equal up to a multiplicative
constant. Hence, the formula (\ref{Eq:UfUfKk}) is proved.

We now consider the case $f(w)=w/\sqrt{1-w^2}$. For this function,
$U_f$ is given by\cite{rf:GRSZ2}
\begin{eqnarray}
 U_f = \exp\left(\,
\frac{1}{2}\,L_2\,\right).
\end{eqnarray}
For $k=-2$, the function $g(w)$ is found to be
\begin{eqnarray}
 g(w)=2\log(1-w^2).
\end{eqnarray}
Then, applying the formula (\ref{Eq:UfUfKk}) to this case, we obtain the
equation
\begin{eqnarray}
\label{Eq:UfUfq} 
U_f^{(-2)}= U'_f\times \exp\left(
\sum_{n=1}^\infty \frac{2}{n}\,q_{2n}
\right),
\end{eqnarray}
where the multiplicative constant can be shown to be 1 by
expanding the both 
sides and using the relations $L_n'=L_n+nq_n+\delta_{n,0}$ and
$L_n^{(k)}=L_n+kq_n+(k^2-3k)\delta_{n,0}/2$.\cite{rf:KT} 
Noting that $(L_n^{(k)})^\dagger=L_{-n}^{(-k)}\ \ (n\neq 0)$, we obtain
the following 
relation by taking the Hermitian conjugate of (\ref{Eq:UfUfq}):
\begin{eqnarray}
\label{Eq:UFKUF}
U'_F = e^{-K(\rho)}\,U_F^{(2)}.
\end{eqnarray}
The operator $K(\rho)$ here is given by (\ref{Eq:Krho}),
and $U_F$ is defined as
\begin{eqnarray}
 U_F = (U_f)^\dagger =\exp\left(\,\frac{1}{2}\,L_{-2}\,\right).
\end{eqnarray}
The function corresponding to $U_F$ is given by $F(w)=\sqrt{w^2+1}$.

Since $\Q^{(2)}$ and $L_n^{(2)}$ commute,
the operator $\Q^{(2)}$ is transformed into $\bQ$
under the similarity transformation with $U'_F$: 
\begin{eqnarray}
 -\frac{1}{4}\,
U'_F\,\Q^{(2)}\,{U'_F}^{-1} 
=-\frac{1}{4}\,e^{-K(\rho)}\,U_F^{(2)}\,\Q^{(2)}\,
(U_F^{(2)})^{-1}\,e^{K(\rho)}=\bQ,
\end{eqnarray}
where we have used the equation (\ref{Eq:KbQK2}).
Consequently, 
from a similar derivation of (\ref{Eq:cohom1}), it follows that if a
state $\ket{\psi}$ satisfies $\bQ\ket{\psi}=0$, 
this state can be written
\begin{eqnarray}
\label{Eq:cohom2}
 \ket{\psi}=U'_F\,(\ket{P}\otimes b_{-2}\ket{0})
+ U'_F\,(\ket{P'}\otimes \ket{0}) +\bQ\ket{\phi}.
\end{eqnarray}
As in the case $a>-1/2$, the difference between the cohomologies
(\ref{Eq:cohom1}) and (\ref{Eq:cohom2}) turns out to be merely
a BRS-exact state, as can be shown by using the relation
(\ref{Eq:UFKUF}). 

For $a=-1/2$, the operator $L(a)$ becomes
\begin{eqnarray}
 \tilde{L}=L(-1/2)=
  \frac{1}{2}L'_0-\frac{1}{4}(L'_2+L'_{-2})+\frac{3}{2}.
\end{eqnarray}
Though the operator $\tilde{L}$ cannot be transformed into $L_0$
under the similarity transformation with $U(s,t,u)$, as seen in the
previous section,
it can be transformed into $L'_2$ under a
transformation with $U'_F$:
\begin{eqnarray}
 \tilde{L}=-\frac{1}{4}\,U'_F\,L'_2\,{U'_F}^{-1}.
\end{eqnarray}

Finally, we consider scattering amplitudes in the theory expanded around
the non-trivial solution.
Because $b_0$ commutes with $L'_n$, the cohomology in the Siegel
gauge is given by
\begin{eqnarray}
 U'_F\,(\ket{P}\otimes b_{-2}\ket{0})
+ U'_F\,(\ket{P'}\otimes \ket{0}).
\end{eqnarray}
These states have ghost number $-1$ or $0$.
If we calculate an amplitude with these as external states, it becomes
zero because its total ghost number differs from that needed
to realize non-zero amplitudes.
Hence, we conclude that
open strings do not appear perturbatively
in the theory expanded around the universal solution at $a=-1/2$ 

\section{Discussion}

In this paper, we have studied open string field theory expanded around
the universal solution under the Siegel gauge condition. In the theory
for $a>-1/2$, we derived physical states and the Feynman rule, and then
we showed that open string amplitudes in the original theory can be
reproduced. In the theory with $a=-1/2$, we proved no open string
theorem, 
which states that no open string exists around the
non-trivial universal solution, because perturbative amplitudes become
zero there. These results provide further evidence for the conjecture
that
universal solutions correspond to a pure gauge or the tachyon vacuum.
Though the function in the solution is restricted to the form given in
(\ref{Eq:ha}), it is possible to extend it to other functions, for
example to the function given in Ref.~\citen{rf:KT} in a straightforward
way. 

For $a>-1/2$, through a string field redefinition,
we can understand the coincidence of the amplitudes of
the original and expanded theories.
The action of the gauge unfixed theory expanded around the solution can
be 
transformed into the original action under the similarity
transformation with the operator $K(h_a)$.\cite{rf:TT2}
Similarly, as found in \S 2.4, 
if we carry out a string field redefinition using the operator $U'(a)$
and change the normalization of the string field,
we can transform the action in the Siegel gauge to the
original action fixed in the same gauge. Consequently, the amplitudes in
the two theories are identical. Here, this scale transformation of
the string field provides the factor absorbed into external states of
the amplitudes. 

It is plausible that the universal solution for $a=-1/2$ is
the tachyon vacuum solution. If this is the case, we should be able to
observe closed 
strings, at least as on-shell states, in the theory expanded around this
solution. This is an open problem, which is important for proving the
equivalence of these solutions and
Sen's conjectures.\cite{rf:Sen-1,rf:Sen-2}
As Drukker pointed out,\cite{rf:Drukker2,rf:Drukker} it should be
possible to reproduce closed string scattering amplitudes 
with the Feynman rule around the non-trivial solution.

\section*{Acknowledgements}
We would like to thank Hiroshi Itoyama for discussions and
encouragement. 

\newpage
\appendix

\section{Generalization of (\ref{Eq:UfUfK})}

In the twisted ghost conformal field theory with a general background
charge,\cite{rf:GRSZ,rf:FMS} the energy momentum tensor is given by
\begin{eqnarray}
 T'(w)=T(w)+(\lambda-2)\,\partial J_{\rm gh}(w),
\end{eqnarray}
where $\lambda$ denotes a real number. 
The theory with $\lambda=1$ corresponds to the case investigated in the
main text.
The mode expansions of the ghost and anti-ghost fields in the twisted
theory are given by
\begin{eqnarray}
 c'(w)=\sum_{n=-\infty}^\infty c_n w^{-n-1+\lambda},\ \ \ 
 b'(w)=\sum_{n=-\infty}^\infty b_n w^{-n-\lambda}.
\end{eqnarray}
Thus, the twisted and untwisted ghost fields have the relations
\begin{eqnarray}
\label{Eq:bclambda}
 c'(w)=w^{\lambda-2}\,c(w),\ \ \ 
 b'(w)=w^{-(\lambda-2)}\,b(w).
\end{eqnarray}
The twisted Virasoro generators can be written in terms of the untwisted
operators as
\begin{eqnarray}
 L_n'=L_n+(2-\lambda)q_n+a_n\,\delta_{n,0},
\end{eqnarray}
where $a_n=-(\lambda-2)(\lambda+1)/2$ if $\lambda=0,\,\pm 1,\,\pm 2,\cdots$.

If the operators $U_f$ and
$U'_f$ implement the conformal mapping $z=f(w)$ in the untwisted and
twisted theories, respectively, the primary fields $\phi(w)$ of
dimension $h$ and $\phi'(w)$ of dimension $h'$ in each theory satisfy
\begin{eqnarray}
\label{Eq:UphiU}
U_f \phi(w)U_f^{-1}=(\partial f(w))^h\,\phi(f(w)),\ \ \ 
U_f' \phi'(w){U_f'}^{-1}=(\partial f(w))^{h'}\,\phi'(f(w)).
\end{eqnarray}
The ghost operators $c(w)$, $b(w)$, $c'(w)$ and $b'(w)$
have dimensions $h=-1$, $h=2$,
$h'=1-\lambda$ and $h'=\lambda$, respectively.  
Using (\ref{Eq:bclambda}) and (\ref{Eq:UphiU}), we find that
\begin{eqnarray}
 U_f e^{K(g)}\,c'(w)\,e^{-K(g)} U_f^{-1} &=&
 w^{\lambda-2}\,U_f e^{K(g)}\,c(w)\,e^{-K(g)} U_f^{-1} \nn
 &=&
 w^{\lambda-2}\,e^{g(w)}\,U_f \,c(w)\,U_f^{-1} \nn
&=&
 w^{\lambda-2}\,e^{g(w)}\,(\partial f(w))^{-1}
 \,c(f(w)) \nn
&=&
 w^{\lambda-2}\,e^{g(w)}\,(\partial f(w))^{-1}
 \,(f(w))^{2-\lambda} c'(f(w)).
\end{eqnarray}
Similarly, we obtain the following equation for the anti-ghost:
\begin{eqnarray}
 U_f e^{K(g)}\,b'(w)\,e^{-K(g)} U_f^{-1} &=&
 w^{-\lambda+2}\,e^{-g(w)}\,(\partial f(w))^2
 \,(f(w))^{\lambda-2} b'(f(w)).
\end{eqnarray}
If these transformation laws coincide with the conformal mapping
$z=f(w)$ in the twisted theory, the function $g(w)$ satisfies
\begin{eqnarray}
&&
 w^{\lambda-2}\,e^{g(w)}\,(\partial f(w))^{-1}
 \,(f(w))^{2-\lambda} = (\partial f(w))^{1-\lambda},\nn
&&
 w^{-\lambda+2}\,e^{-g(w)}\,(\partial f(w))^2
 \,(f(w))^{\lambda-2} 
= (\partial f(w))^\lambda.
\end{eqnarray}
In this case, $g(w)$ is given by
\begin{eqnarray}
 g(w)=\log \left(\frac{w \partial f(w)}{f(w)}\right)^{2-\lambda}.
\end{eqnarray}
Because the operator $U'_f$ that induces the conformal mapping $z=f(w)$
can be uniquely determined up to a multiplicative constant, we obtain
the formula 
\begin{eqnarray}
 U_f'={\rm const}\times
U_f\times \exp
K\left(\log\left(
\frac{w\,\partial f(w)}{f(w)}
\right)^{2-\lambda}\right).
\end{eqnarray}
If we set $\lambda=1$, this formula reduces to the equation
(\ref{Eq:UfUfK}). 


\end{document}